\documentclass[prd,preprintnumbers,nofootinbib,tightenlines,superscriptaddress]{revtex4}
\usepackage{graphicx,epstopdf}
\usepackage{amsmath,amssymb,graphicx}
\usepackage[usenames, dvipsnames]{color}
\usepackage{slashed}
\usepackage{times}
\usepackage{latexsym}
\usepackage{floatrow}
\usepackage{longtable}
\usepackage[utf8, latin9]{inputenc}
\usepackage{hyperref}

\begin{document}

\title{Density spikes near black holes in self-interacting dark matter halos and indirect detection constraints}

\author{Gerardo Alvarez}
\email[]{gerardo.alvarez@email.ucr.edu}
\affiliation{Department of Physics and Astronomy, University of California, Riverside, California 92521, USA}
\author{Hai-Bo Yu}
\email[]{hai-bo.yu@ucr.edu}
\affiliation{Department of Physics and Astronomy, University of California, Riverside, California 92521, USA}

\date{\today}

\begin{abstract}
Self-interacting dark matter (SIDM) naturally gives rise to a cored isothermal density profile, which is favored in observations of many dwarf galaxies. The dark matter distribution in the presence of a central black hole in an isothermal halo develops a density spike with a power law of $r^{-7/4}$, which is shallower than $r^{-7/3}$ as expected for collisionless dark matter (CDM). Thus, indirect detection constraints on dark matter annihilations from the density spike could be relaxed in SIDM. Taking the most dense satellite galaxy of the Milky Way Draco as an example, we derive upper limits on the annihilation cross section and the black hole mass for both SIDM and CDM halos. For the former case, Draco could host an intermediate mass black hole even if dark matter is composed of thermal relics. We further explore the constraints from the Milky Way and M87, which host supermassive black holes, and show the upper limits on the annihilation cross section are significantly weakened in SIDM. Our results also indicate that the Event Horizon Telescope could provide a unique test of SIDM spikes.

\end{abstract}


\maketitle

\section{Introduction}
\label{sec:intro}

It is well established that dark matter makes up about $85\%$ of the mass in the Universe~\cite{Ade:2015xua}. Indirect searches of high-energy standard model particles originating from dark matter annihilations or decays provide an important way of understanding its nature. Of central importance to the indirect detection search is the distribution of dark matter within galactic halos. In the prevailing scenario of dark matter, it is composed of collisionless thermal relics. Numerical simulations show that collisionless dark matter (CDM) typically produces a Navarro-Frenk-White (NFW) density profile in the halo~\cite{Navarro:1995iw,Navarro:1996gj}, which has a characteristic feature of $\rho(r)\propto r^{-1}$ towards the center~\cite{Dubinski:1991bm}. It has also been established that virtually all large galaxies host central supermassive black holes, see, e.g.~\cite{Kormendy:2013dxa}. The presence of a central black hole could alter the dark matter density profile in the inner halo, and for the standard NFW profile a density spike, $\rho(r)\propto r^{-7/3}$, could form near a black hole that grew adiabatically~\cite{Gondolo:1999ef}. The high density of the spike could boost the dark matter annihilation rate. For example, the Milky Way hosts a central supermassive black hole with mass $\sim4\times10^6~{\rm M_\odot}$~\cite{Gillessen:2008qv,Ghez:2008ms}. Dark matter annihilations could produce very bright sharp signals that may be visible as a point source in the galactic center, see, e.g.,~\cite{Gondolo:1999ef,Hooper:2002ru,Aloisio:2004hy,Fields:2014pia,Belikov:2016fwv,Sandick:2016zeg,Chiang:2019zjj}. Refs.~\cite{Lacroix:2015lxa,Lacroix:2016qpq} show that observations from the M87 galaxy have excluded thermal relic dark matter in the presence of a CDM spike near its $6.5\times10^9~{\rm M_\odot}$ black hole. 

It's less known observationally whether small dwarf galaxies, like satellite galaxies of the Milky Way, may host central black holes with intermediate masses. Interestingly, one could derive constraints on the central black hole mass in the satellites using dark matter indirect detection observations. For example, for the thermal relic scenario with an $s$-wave annihilation cross section, the central black hole mass cannot be higher than $\sim10^2\textup{--}10^3~{\rm M_\odot}$ in Draco~\cite{Wanders:2014xia}, the most dense satellite of the Milky Way, otherwise the flux of dark matter annihilation signals would surpass the upper limit from Fermi-LAT gamma-ray observations due to the presence of the density spike induced by the black hole. These limits could be relaxed if the power law of the spike is more mild. This could occur if the black hole grows away from the center of the halo or it does not grow adiabatically from a seed, but being brought in by mergers~\cite{Ullio:2001fb}. In addition, mergers of black holes in the centers of the progenitor halos could erase the density spike~\cite{Merritt:2002vj}. Gravitational scatterings between stars and dark matter particles could also kinetically heat up the spike and reduce its density~\cite{Gnedin:2003rj,Merritt:2003qk}.

In this work, we study indirect detection constraints on dark matter spikes in self-interacting dark matter (SIDM), see~\cite{Tulin:2017ara} for a recent review. In this scenario, dark matter has strong self-interactions that can thermalize the inner halo over cosmological timescales~\cite{Dave:2000ar,Vogelsberger:2012ku,Rocha:2012jg,Peter:2012jh,Zavala:2012us,Vogelsberger:2015gpr,Fischer:2020uxh}. Recent studies show that SIDM is favored for explaining diverse dark matter distributions over a wide range of galactic systems~\cite{Kaplinghat:2015aga,Kamada:2016euw,Creasey:2016jaq,Ren:2018jpt,Kaplinghat:2019svz,Sameie:2019zfo,Kahlhoefer:2019oyt,Yang:2020iya,Sagunski:2020spe,Andrade:2020lqq}, implying that the inner region of dark halos might be indeed thermalized. Ref.~\cite{Shapiro:2014oha} uses a conduction fluid model and derives density profiles of SIDM particles bound to a black hole. For a Coulomb-like self-interaction, a central black hole can induce a density spike of $\rho(r)\propto r^{-7/4}$, which is shallower than the CDM one $\rho(r)\propto r^{-7/3}$. We apply these results to Draco, the Milky Way and M87, and derive upper limits on the annihilation cross section. We will show in SIDM Draco could contain central black holes with intermediate masses $\sim600~{\rm M_\odot}$ as expected in the black hole-host galaxy relation~\cite{Tremaine:2002js}, while the halo is still composed of thermal relic dark matter. And the upper limits can be weakened by factors of $\sim10^7$ and $10^3$ for the Milky Way and M87, respectively. We will also show that M87 could be a promising target for probing SIDM spikes with data from the Event Horizon Telescope (EHT). 

The rest of the paper is organized as follows. In Sec. II, we review density profiles of SIDM spikes near a black hole. In Sec. III, we derive Fermi-LAT constraints on the annihilation cross section and the black hole mass for Draco. In Sec. IV, we discuss implications for the Milky Way and M87. We conclude in Sec. V.

\section{SIDM density spikes near a black hole}

We assume dark matter particles in the inner halo follow an isotropic, quasi-equilibrium distribution due to dark matter interactions. A Schwarzschild black hole is located at the center of the halo with a mass of $M_\bullet$, which is much smaller than the total halo mass, but larger than the mass of bound SIDM particles in the spike. Ref.~\cite{Shapiro:2014oha} uses a conduction fluid model and derives density profiles of SIDM particles bound to the black hole, which depend on the form of the self-scattering cross section $\sigma$. Consider the parameterization $\sigma=\sigma_0(v/v_0)^{-a}$, where $\sigma_0$ is the normalization factor, $a$ characterizes the velocity dependence, $v$ is 1D velocity dispersion of the particles in the spike and $v_0$ is that of outside, the density follows a power law of  
\begin{equation}
\rho\propto r^{-(3+a)/4}
\label{eq:sidmspike}
\end{equation}
for $r\lesssim r_{\rm bh}=GM_\bullet/v^2_0$, and the corresponding velocity dispersion scales as $v\propto r^{-1/2}$~\cite{Shapiro:2014oha}. With the power law solution, it's easy to see that the energy flux transported out of the spike due to the self-interactions $L$ is independent of radius, as $L\sim N E/t_r$, where $N\sim\rho r^3$ is the number of bound particles per shell, $E\sim v^2$ energy per particle and $t_r=1/(\sigma\rho v)$ relaxation time. This is the condition to have a steady state near a black hole~\cite{Shapiro:1976}. 

We see that the spike density profile becomes steeper as $a$ increases. Since the cross section is more suppressed in the spike for higher $a$, the transport rate becomes smaller accordingly, resulting in a higher density. The velocity dependence of $\sigma$ is related to particle physics realizations of SIDM. For example, a scalar dark matter candidate could have a self-coupling that leads to a constant cross section and $a=0$ over all scales. More generally, there exists a scalar or vector force mediator with a mass of $m_\phi$. When $m_\chi v_0> m_\phi$, the self-scattering is Coulomb-like, i.e., $\sigma\propto v^{-4}_0$ and $a=4$. In the opposite limit, it's point-like and $a=0$. And in the resonant regime for attractive interactions, $\sigma\propto v^{-2}_0$ and $a=2$. For a given set of mass parameters, $a$ may vary as well. Consider the best-fit model in~\cite{Kaplinghat:2015aga,Huo:2017vef}, where $m_\chi\sim10^{3}m_\phi$. For clusters, $v_0\sim10^3~{\rm km/s}$, the model predicts $a=4$ in both spike and its surrounding regions. For a dwarf halo with $v_0\sim100~{\rm km/s}$, the self-scattering is point-like in the bulk of the halo, but becomes Coulomb-like towards the inner spike as $v$ increases as $r^{-1/2}$. In addition, black hole and halo masses are correlated. Thus observations of SMBHs over different mass scales may provide a unique probe of SIDM models.

As discussed above, the steepest spike density profile predicted in SIDM is $\rho\propto r^{-7/4}$ for $a=4$, which is slightly shallower than the one predicted in CDM if the hole grows adiabatically, i.e., $\rho\propto r^{-7/3}$. We will show that the small difference in the logarithmic density slope could lead to significantly different constraints from indirect detection as the signal strength is $\propto\rho^2(r)$. We note that CDM could have a spike profile of $\rho\propto r^{-3/2}$~\cite{Gondolo:1999ef}. Frequent gravitational scatterings between stars near the black hole and CDM particles could drive the latter to follow an isothermal distribution~\cite{Gnedin:2003rj,Merritt:2003qk}. This effect could be important for the Milky Way, but it's negligible for Draco and M87, as we will discuss later. 

To study implications of the density spikes on indirect detection constraints, we need to further specify inner boundary conditions. For SIDM, we extend the spike profile in Eq.~\ref{eq:sidmspike} to $r=2r_{\bullet}$~\cite{Ferrer:2017xwm,Shapiro:2014oha}, where $r_{\bullet}=2GM_\bullet/c^2$ is the Schwarzschild radius, and set $\rho(r)=0$ for $r\leq2 r_{\bullet}$. For CDM, we consider annihilation radius $r_{\rm ann}$ that is calculated iteratively as $\rho (r_{\rm{ann}}) = m_\chi/\left < \sigma_{\rm{ann}} v_{\rm rel} \right > t_{\rm age}$~\cite{Vasiliev:2007vh,Shapiro:2016ypb}, where $\left < \sigma_{\rm{ann}} v_{\rm rel} \ \right >$ is the thermally averaged annihilation cross section and $t_{\rm age}=10~{\rm Gyr}$ is the age of the system. The CDM spike density saturates to $\rho(r_{\rm ann})$ at the annihilation radius, and we further set $\rho(r)=0$ for $r\leq 2r_\bullet$~\cite{Ferrer:2017xwm}. Note in SIDM, dark matter self-interactions could wash out the annihilation plateau~\cite{Shapiro:2016ypb}.

\section{Applications to Draco}

\begin{figure}[t!]
\includegraphics[scale=0.35]{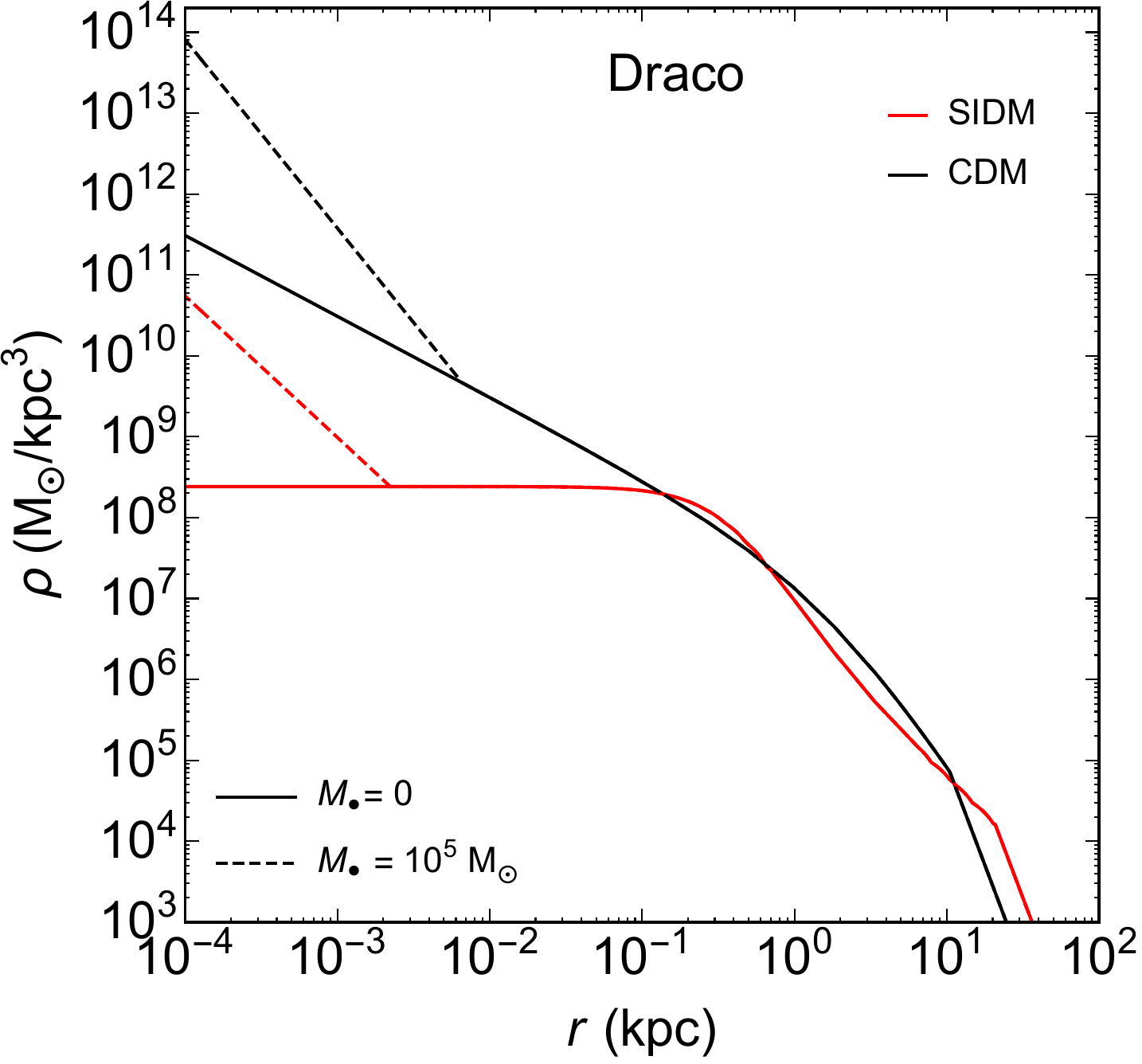}\;\;
\includegraphics[scale=0.35]{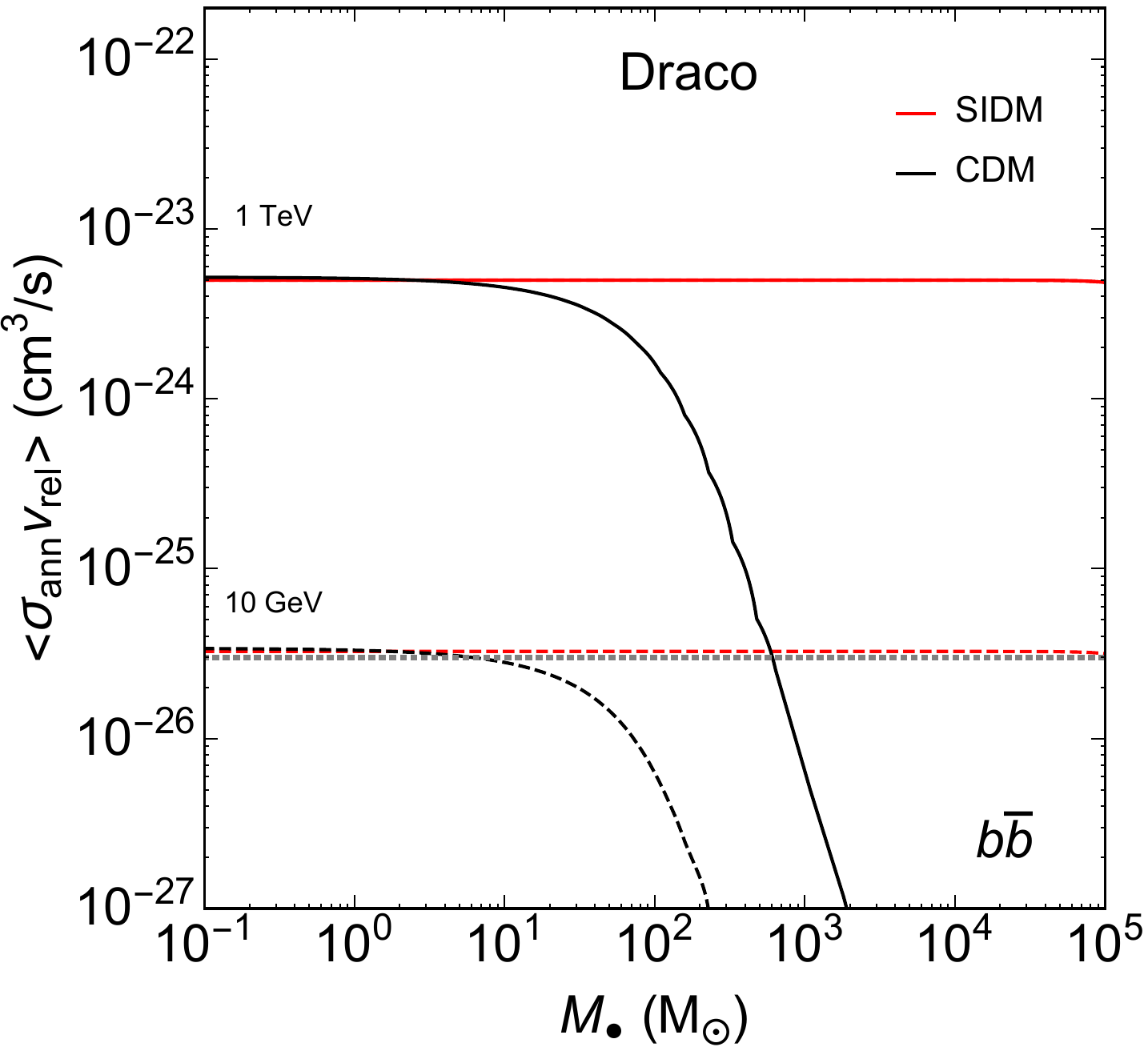}
\caption{{\it Left:} dark matter halo density profiles for SIDM (red) and CDM (black) halo models for the Milky Way satellite galaxy Draco without (solid) and with (dashed) a central black hole. For the latter case, the black hole mass is assumed to be $10^{5}~{\rm M_{\odot}}$ and the density spike follows a power law of $r^{-7/4}$ and $r^{-7/3}$ for the SIDM and CDM halos, respectively. A steep cutoff at around $10~{\rm kpc}$ is due to tidal stripping. {\it Right:} upper limits on the annihilation cross section vs. the central black hole mass for the SIDM (red) and CDM (black) halos with density spikes, based on Fermi-LAT gamma-ray observations of Draco. The dark matter mass is $10~{\rm GeV}$ (dashed) and $1~{\rm TeV}$ (solid). The horizontal line denotes the canonical thermal annihilation cross section $\left<\sigma_{\rm ann}v_{\rm rel}\right>=3\times10^{-26}~{\rm cm^3/s}$ (dotted gray).}
\label{fig:draco}
\end{figure}

The presence of dark matter spikes could significantly boost indirect detection signals. In turn, if we assume dark matter is made of thermal relics, we can derive constraints on the mass of central black holes of galaxies using results from dark matter indirect searches. Ref.~\cite{Wanders:2014xia} considers Draco, the most dense satellite of the Milky Way, and shows its black hole mass cannot be larger than $10^2\textup{--}10^3~{\rm M_{\odot}}$ for the dark matter mass in the range $100~{\rm GeV}\textup{--}1~{\rm TeV}$, based on the Fermi-LAT gamma-ray data. 

Since the SIDM halo model predicts a shallower density spike than the CDM one, we expect that the upper limits on $M_\bullet$ for Draco could be relaxed accordingly. To see the difference, we first consider dark matter density profiles for Draco without a black hole. Ref.~\cite{Kaplinghat:2019svz} fits the line-of-sight stellar velocity dispersion of Draco with an NFW profile and finds the best-fit values of the scale density and radius are $\rho_s\approx1.68\times10^{7}~{\rm M_\odot/kpc^3}$ and $r_s\approx1.94~{\rm kpc}$, respectively. In addition, it also considers a cored isothermal density profile following the solution to the Jeans equation $v^2_0\nabla^2\ln\rho=-4\pi G\rho$ with the boundary conditions $\rho(0)=\rho_0$ and $\rho'(0)=0$, and finds the best fit values $\rho_0\approx2.55\times10^8~{\rm M_\odot/kpc^3}$ and $v_0\approx13.88~{\rm km/s}$. This isothermal profile was first proposed to model dark matter distributions in an inner SIDM halo~\cite{Kaplinghat:2015aga,Kaplinghat:2013xca} and it agrees with N-body simulations remarkably well~\cite{Robertson:2017mgj,Ren:2018jpt,Robertson:2020pxj}. The left panel of Fig.~\ref{fig:draco} (solid) shows the dark matter density profiles for Draco inferred from fitting to stellar kinematics as in~\cite{Kaplinghat:2019svz}, where we have extrapolated them for $r\gtrsim10~{\rm kpc}$ using a power law of $r^{-5}$ to account for tidal stripping.

We use $\rho_{\rm{spike}}(r) = \rho(r_{\rm spike}) \left ( r_{\rm{spike}} / r \right )^{\gamma}$ to model the spike density profile, where  $r_{\rm spike}$ is the spike radius, and $\gamma=7/4$ and $7/3$ for SIDM and CDM halos, respectively. For SIDM, we set the spike radius to be the radius of influence calculated as $r_{\rm bh}=G M_{\bullet} / v_{0}^{2}$, where $v_0$ is the 1D velocity dispersion outside of the spike. It's important to note that $v_0$ is a constant over the radius for an SIDM halo and the calculation of $r_{\rm bh}$ is self-consistent. Taking $M_{\bullet}=10^{5}~{\rm M_\odot}$ as an example and $v_{0} \approx 14~{\rm km/s}$ for Draco~\cite{Kaplinghat:2019svz}, we find $r_{\rm spike}\approx2.2~{\rm pc}$ and $\rho(r_{\rm spike})\approx2.4\times10^8~{\rm M_\odot/kpc^3}$ as shown in the left panel of Fig.~\ref{fig:draco} (dashed red). For CDM, $v_0$ is not a constant and it depends on radius. In this case, we follow~\cite{Merritt:2003qk} and adopt a practical definition of $r_{\rm bh}$ through the condition
\begin{equation}
\label{eq:mass}
4\pi \int_{0}^{r_{\rm bh}}dr r^{2} \rho(r) = 2 {M_{\bullet}},
\end{equation}
and the CDM spike radius is given by $r_{\rm{spike}}\approx0.2 r_{\rm{bh}}$~\cite{Merritt:2003qc}. For $M_{\bullet}=10^{5}~{\rm M_\odot}$, we have $r_{\rm bh}\approx33~{\rm pc}$, hence $r_{\rm spike}\approx6.6~{\rm pc}$ and $\rho(r_{\rm spike})\approx4.6\times10^9~{\rm M_\odot/kpc^3}$; see the left panel of Fig.~\ref{fig:draco} (dashed black). Note the annihilation radius is much smaller than the spike radius and we do not show it in the figure. For instance, consider $\left<\sigma_{\rm ann}v_{\rm rel}\right>=3\times10^{-26}~{\rm cm^3/s}$ and $m_\chi=100~{\rm GeV}$, we find $r_{\rm ann}\approx10^{-4}~{\rm pc}$ for the CDM halo.

We consider dark matter annihilations to ${b\bar{b}}$ states, which further produce gamma-ray signals. The differential flux from the contribution of the smooth halo component can be calculated as 
\begin{equation}
\label{eq:halo}
\frac{d \Phi_{\rm{halo}}}{dE} = \frac{1}{2} \frac{\left <  \sigma_{\rm{ann}} v_{\rm rel} \right >}{4 \pi m_{\chi}^{2}} \frac{d N}{ dE} \bar{J}
\end{equation}
where $dN/dE$ is the photon spectrum and $\bar{J}$ is the angular integrated $J$ factor given by
\begin{equation}
\bar{J}=2\pi\int^{\theta_{\rm{max}}}_{0} d\theta\sin\theta J(\theta)= 2\pi\int^{\theta_{\rm{max}}}_{0}d\theta \sin\theta \int_{l.o.s} ds \rho^{2}[r(\theta,\rm{s})].
\label{eq:jfactor}
\end{equation}
To perform the integral along the line of sight direction, we write $r(s,\theta) = \sqrt{D^2 + s^{2} - 2 s D  \cos(\theta)}$, where $D\approx76~{\rm kpc}$ is the distance from Earth to Draco, and we set $\theta_{\rm max}=0.5^{\rm o}$, corresponding to a solid angle of $2.4\times10^{-4}~{\rm sr}$. For the SIDM and CDM halos of Draco shown in the left panel Fig.~\ref{fig:draco} (solid), we find $\bar{J}\approx5.0\times10^{18}~{\rm GeV^2/cm^5}$ and $4.8\times10^{18}~{\rm GeV^2/cm^5}$, respectively. We see that although the two halo models have very different inner density profiles, their $\bar{J}$ factors are similar. 

For the contribution from the density spike, we have
\begin{equation}
\label{eq:spike}
\frac{d\Phi_{\rm{spike}}}{dE} = \frac{1}{2} \frac{\left < \sigma_{\rm{ann}} v_{\rm rel} \right >}{m_{\chi}^{2}D^{2}} \frac{dN}{dE} Q,
\end{equation}
where the $Q$ factor is calculated as
\begin{equation}
\label{eq:qfactor}
Q=\int_{r_{\rm min}}^{{r_{\rm spike}}} dr r^{2} \rho_{\rm{spike}}^{2}(r)=
\begin{cases}
2\rho^2(r_{\rm spike})r^3_{\rm spike}\left(\frac{r_{\rm spike}}{2r_{\bullet}}\right)^{1/2},\, {\rm SIDM} \\
\frac{3}{5}\rho^2(r_{\rm spike})r^3_{\rm ann}\left(\frac{r_{\rm spike}}{r_{\rm ann}}\right)^{14/3},\, {\rm CDM} 
\end{cases}
\end{equation}
where $r_{\rm min}$ is $2 r_{\bullet}=4 { G} M_{\bullet} / c^{2}$ for SIDM and the annihilation radius $r_{\rm ann}$ for CDM. The annihilation radius can be calculated as $r_{\rm ann}=[\left<\sigma_{\rm ann}v_{\rm rel}\right>t_{\rm age}\rho(r_{\rm spike})/m_\chi]^{3/7}r_{\rm spike}$. For simplicity, we neglect contributions from the annihilation plateau, which could underestimate the CDM limits on $\left < \sigma_{\rm{ann}} v_{\rm rel} \right >$ by $\sim30\%$ for Draco. In addition, we have assumed the condition $r_{\rm min} \ll r_{\rm spike}$ in Eq.~\ref{eq:qfactor}, which is valid for the system we consider. For a given dark matter mass, black hole mass and annihilation cross section, we can calculate the expected signal flux by integrating Eqs.~\ref{eq:halo} and~\ref{eq:spike} for $100~{\rm MeV}\leq E\leq100~{\rm GeV}$, the energy range of the Fermi-LAT gamma-ray space telescope. We take the photon spectrum $dN/dE$ from~\cite{Cirelli:2010xx,Amoroso:2018qga}, and obtain the total flux as $\Phi_{\rm{total}}=\Phi_{\rm{halo}}+\Phi_{\rm{spike}}$. For $b\bar{b}$ final states, the upper limit on the gamma-ray flux is $\Phi_{\rm upper}\approx(62\textup{--}5.8)\times10^{-11}~{\rm cm^{-2}s^{-1} }$ for $m_\chi=10~{\rm GeV}\textup{--}1~{\rm TeV}$~\cite{Wanders:2014xia}, based on Fermi-LAT data on Draco. We vary the black hole mass and the dark matter mass, and derive upper limits on the annihilation cross section for the SIDM and CDM halo models of Draco. 

The right panel of Fig.~\ref{fig:draco} shows the upper limits on the annihilation cross section vs. the black hole mass for the SIDM (red) and CDM (black) halos, where we consider the dark matter mass $m_\chi=10~{\rm GeV}$ (dashed) and $1~{\rm TeV}$ (solid). The gray horizontal line denotes the canonical thermal cross section $\left<\sigma_{\rm ann}v_{\rm rel}\right>=3\times10^{-26}~{\rm cm^3/s}$. For the CDM halo, the thermal relic dark matter is excluded for $m_\chi=10~{\rm GeV}$ and $1~{\rm TeV}$ if $M_\bullet\gtrsim10~{\rm M_\odot}$ and $10^3~{\rm M_\odot}$, respectively. For the SIDM halo, the upper limits on the annihilation cross section are essentially independent of the black hole mass, and the constraints are significantly relaxed. Since both halo models have similar $\bar{J}$ factors for the smooth component, the difference in the $M_\bullet$ bounds is caused by their different spike profiles. Observationally, it's unknown whether Draco has a massive central black hole. If we extrapolate the black hole-host galaxy relation~\cite{Tremaine:2002js} to Draco, it could host a black hole with $M_\bullet\sim600~{\rm M_\odot}$. In this case, SIDM could be composed of thermal relics, but CDM could not be for $m_\chi\lesssim1~{\rm TeV}$. In the limit where the spike is negligible, our analysis shows thermal relic dark matter is allowed for both halo models with $m_\chi\sim10~{\rm GeV}$. Ref.~\cite{Wanders:2014xia} finds a stronger limit of $m_\chi\gtrsim30~{\rm GeV}$ for a CDM halo. This is because it considers a density profile with $\bar{J}\approx1.20\times10^{19}~{\rm GeV^2/cm^5}$, which is a factor of $2.5$ higher than our case. 

SIDM predicts a weaker density spike near a central black hole. For the satellite galaxies like Draco, the presence of such a spike does not strengthen constraints on the SIDM annihilation cross section unless the black hole mass is much larger than $10^7~{\rm M_\odot}$, which is impossible for those systems given their small masses. As estimated in~\cite{Sameie:2019zfo}, the halo mass of Draco is about $2\times10^8~{\rm M_\odot}$ with a $4\times10^{9}~{\rm M_\odot}$ progenitor falling into the tidal field of the Milky Way. In this work, we focus on dark matter annihilations to $b\bar{b}$, as it is one of the most studied channels in dark matter indirect detection, but it's straightforward to extend to other channels as well. In addition, we could interpret our results in terms of a specific particle physics model of SIDM, combining with other constraints, see. e.g.,~\cite{Tulin:2013teo,Kaplinghat:2015gha,Bringmann:2016din,Cirelli:2016rnw,Chu:2016pew,Kamada:2018kmi,Chu:2018fzy,Bernal:2019uqr,Kang:2020yul}. It is also interesting to note that Draco was considered as a challenging case for SIDM~\cite{Valli:2017ktb,Read:2018pft} as it has the highest dark matter content among the Milky Way satellites, but both dark matter self-interactions and tidal interactions are commonly expected to produce a shallow density core for a satellite galaxy. Recent works show the interplay of the two effects could actually lead to an opposite consequence, resulting in a high central density, due to the onset of SIDM core collapse~\cite{Nishikawa:2019lsc,Sameie:2019zfo,Kahlhoefer:2019oyt,Correa:2020qam,Turner:2020vlf}. Ref.~\cite{Sameie:2019zfo} uses N-body simulations and demonstrates that the isothermal density profile of Draco shown in the left panel of Fig.~\ref{fig:draco} can be produced in SIDM. 

\section{Implications for the Milky Way and M87}

\begin{figure}[t!]
\includegraphics[scale=0.45]{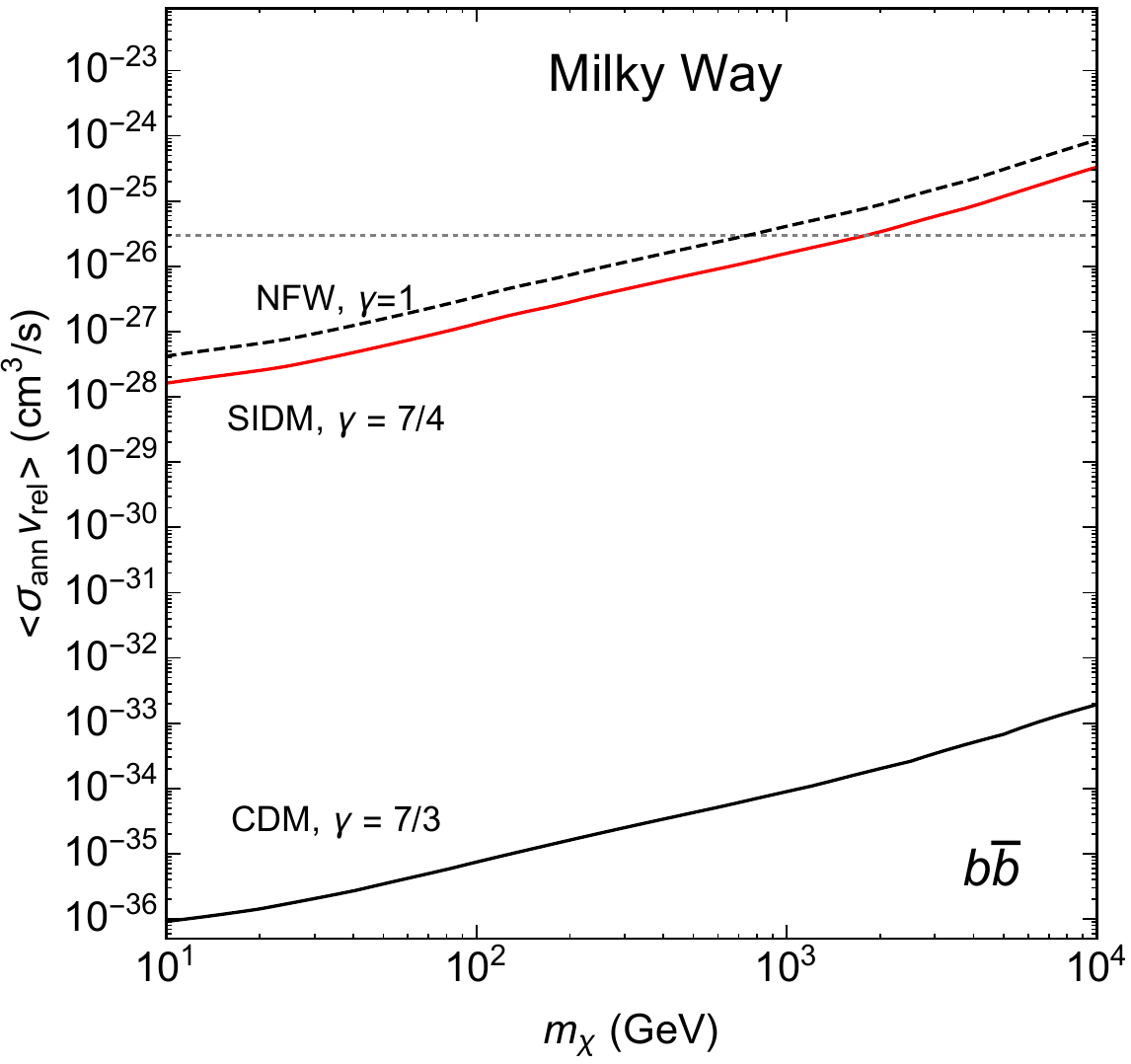}\;\;
\includegraphics[scale=0.45]{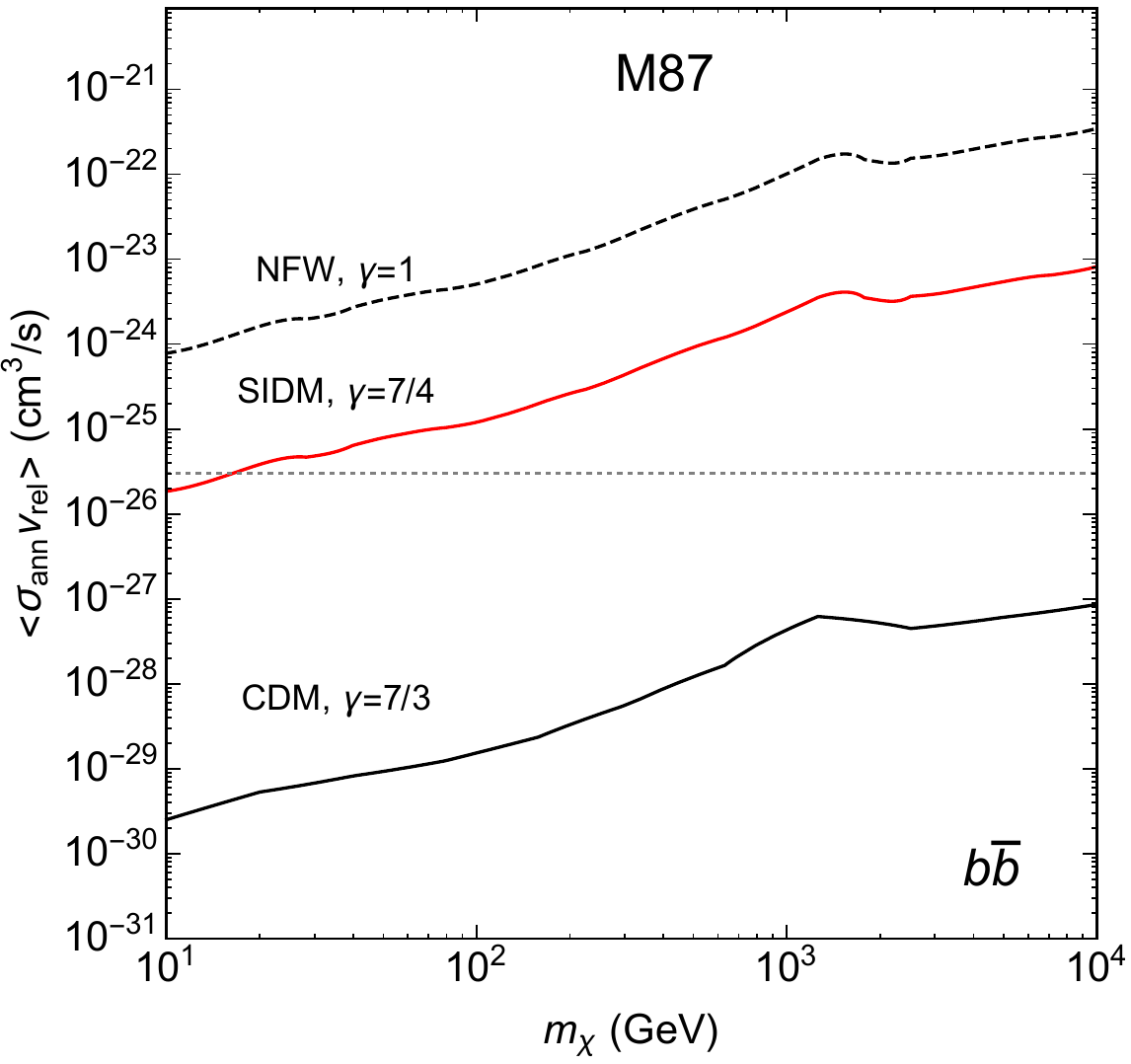}
\caption{{\it Left:} upper limits on the annihilation cross section for the Milky Way with a central black hole mass of $M_{\bullet} = 4\times10^{6}~{\rm M_{\odot}}$ in the presence of SIDM (solid red) and CDM (solid black) spikes, compared to the case assuming a pure NFW halo (dashed black) from~\cite{Abazajian:2020tww}. {\it Right:} similar to the left panel, but for the M87 galaxy with $M_{\bullet} = 6.5 \times 10^{9}~{\rm M_{\odot}}$. The NFW limits are from~\cite{Lacroix:2015lxa}. For both panels, the horizontal line denotes the canonical thermal annihilation cross section $\left<\sigma_{\rm ann}v_{\rm rel}\right>=3\times10^{-26}~{\rm cm^3/s}$ (dotted gray).}
\label{fig:MW_M87}
\end{figure}

We consider the Milky Way, which hosts a central black hole mass with $M_{\bullet}\sim4\times10^6~{\rm M_\odot}$~\cite{Gillessen:2008qv,Ghez:2008ms}. Such a massive black hole could significantly enhance the spike density and boost indirect detection signals accordingly. Taking Fermi-LAT observations of the Galactic Center in gamma rays, Ref. ~\cite{Abazajian:2020tww} derives stringent upper limits on the annihilation cross section for a pure NFW profile and finds $\left<\sigma_{\rm ann} v_{\rm rel}\right>\lesssim4\times10^{-28}\textup{--}9\times10^{-25}~{\rm cm^3/s}$ for $m_\chi\approx10~{\rm GeV}\textup{--}10~{\rm TeV}$ and the $b\bar{b}$ channel. We recast these limits to constrain $\left<\sigma_{\rm ann} v_{\rm rel}\right>$ in the presence of a density spike near the black hole for in both SIDM and CDM. In our study, we demand that the predicted signal flux with a spike should not exceed the one expected from a pure NFW halo as in~\cite{Abazajian:2020tww}.

We first calculate the normalization factor for the flux. We assume an NFW density profile for the Milky Way halo with $r_{s} = 26~{\rm kpc}$ and $\rho_{s}=4.1\times10^{6}~{\rm M_{\odot}/ kpc^{3}}$, consistent with the mean values used in~\cite{Abazajian:2020tww}. The ${\bar J}$ factor takes the form $\bar{J}=\int d\ell db \int_{l.o.s.} ds \rho^{2}[r(\ell,b,s)]$ where $b$ and $\ell$ are Galactic latitude and longitude, respectively, and $r(\ell,b,s)=\sqrt{{ D}^{2}+s^{2}-2 { s D}  {\cos}(\ell){\cos}(b)}$ with $D=8.25~{\rm kpc}$. We integrate both $b$ and $\ell$ from $-20^{\rm o}$ to $20^{\rm o}$, in accord with the signal region in~\cite{Abazajian:2020tww}, and find $\bar{J} = 2.9\times10^{22} \ {\rm GeV^{2}/cm^{5}}$. For a given dark matter mass, we take the corresponding upper limit on $\left<\sigma_{\rm ann} v_{\rm rel}\right>$ from~\cite{Abazajian:2020tww} and convert it into an upper limit on the differential gamma-ray flux using Eq.~\ref{eq:halo}.

Unlike Draco, the stellar mass dominates the inner regions of the Milky Way and SIDM thermalization with a deep baryonic potential could lead to a high density with a negligible core size~\cite{Kaplinghat:2013xca}, as dense as an NFW halo; see~\cite{Sameie:2018chj} for an example. Thus we can approximate the Milky Way halo in SIDM with the NFW profile for $r\geq r_{\rm spike}$ and match it with the spike $\rho(r)\propto r^{-7/4}$ for $2 r_\bullet<r<r_{\rm spike}=r_{\rm bh}=GM/v^2_0$, where we estimate the 1D velocity dispersion as $v_{0} = v_{\rm max}/ \sqrt{3}$ and $v_{\rm max}= 1.64  r_{ s} \sqrt{{G} \rho_{ s}}$~\cite{Kaplinghat:2015aga}. The presence of the stellar mass could further increase $v_0$, resulting in smaller $r_{\rm bh}$. Thus our estimation of $v_0$ could lead to a conservative limit on $\left<\sigma_{\rm ann} v_{\rm rel}\right>$. We then calculate the $Q$ factor given in Eq.~\ref{eq:qfactor} for the SIDM spike. For CDM, we follow the procedure discussed for Draco to calculate the spike density and radius. For $2r_\bullet<r<r_{\rm spike}$, we take the inner profile to be the geometric mean between the annihilation density and the spike density, i.e., $\rho_{\rm spike}(r) \rho (r_{\rm ann}) /[\rho (r_{\rm spike}) + \rho(r_{\rm ann})]$; for $r\leq2r_\bullet$, $\rho(r)=0$~\cite{Gondolo:1999ef}. For the Milky Way, we find $r_{\rm spike}=1.7~{\rm pc}$ and $20~{\rm pc}$ for SIDM and CDM spikes, respectively.

The left panel of Fig. ~\ref{fig:MW_M87} shows the upper limits on the annihilation cross section from the Milky Way after taking into account SIDM (solid red) and CDM (solid black) spikes, compared to the ones assuming a pure NFW halo (dashed black) from~\cite{Abazajian:2020tww}. For SIDM, the presence of the SMBH has a mild effect and only increases the upper limits on $\left<\sigma_{\rm ann} v_{\rm rel}\right>$ by order unity. In the presence of a CDM spike, the limits are a factor of $4\times10^8$ stronger, compared to the pure NFW case. Thus the thermal relic scenario is ruled out for the entire mass range in CDM, while it's allowed for $m_\chi\gtrsim2~{\rm TeV}$ in SIDM. For a CDM spike of $\rho(r)\propto r^{-3/2}$, caused by dynamical heating by stars, we find the limits are similar to the SIDM ones.

Lastly, we consider the supergiant elliptical galaxy M87, which hosts a central black hole mass with $M_{\bullet}\approx6.5\times10^9~{\rm M_\odot}$~\cite{Gebhardt:2009cr,Akiyama:2019eap}. Ref.~\cite{Lacroix:2015lxa} assumes a spike with $\rho(r)\propto r^{-7/3}$ for M87, and derives upper limits on the annihilation cross section to $b\bar{b}$ as $\left<\sigma_{\rm ann} v_{\rm rel}\right>\lesssim6\times10^{-30}\textup{--}10^{-26}~{\rm cm^3/s}$ for the dark matter mass in a range of $m_\chi\approx10~{\rm GeV}\textup{--}100~{\rm TeV}$, a factor of $10^6$ stronger compared to a pure NFW halo. Thus, for CDM, thermal relic dark matter has been excluded for the entire mass range for M87. Compared to the Milky Way, M87 is dynamically young and the CDM spike is expected to survey as gravitational heating is insufficient~\cite{Lacroix:2015lxa}. 

We recast the limits to the case with an SIDM spike, using the approach for the Milky Way discussed previously. For the M87 halo, we take the NFW parameters $r_{s} = 20~{\rm kpc}$ and $\rho_{s}= 6.6\times10^{6}~{\rm M_{\odot}/ kpc^{3}}$ following~\cite{Lacroix:2015lxa}. The right panel of Fig. ~\ref{fig:MW_M87} shows the upper limits on the annihilation cross section from M87 after taking into account SIDM (solid red) and CDM (solid black) spikes. Our CDM limits are stronger than those in~\cite{Lacroix:2015lxa} by a factor of $2.6$, as we take the inner boundary to be $2r_{\bullet}$~\cite{Ferrer:2017xwm}, smaller than theirs $4r_{\bullet}$. For SIDM, the presence of the SMBH increases the upper limits on $\left<\sigma_{\rm ann} v_{\rm rel}\right>$ by a factor of $40$, compared to the case with a pure NFW halo (dashed black). For a CDM spike, the limits are a factor of $4\times10^5$ stronger. Thermal relic scenario is excluded in CDM, while it's allowed for $m_\chi\gtrsim20~{\rm GeV}$ in SIDM. This lower bound is weaker than the Milky Way one, $m_\chi\gtrsim2~{\rm TeV}$. 

Another exciting aspect about the M87 target is that the EHT could resolve the dark matter density profile near the hole, due to its unprecedented angular resolution. Ref.~\cite{Lacroix:2016qpq} shows the EHT is sensitive to synchrotron emission induced by dark matter annihilations and the radiation from the annihilations could further enhance the photon ring around the shadow of the black hole. For a CDM spike with the $b\bar{b}$ channel, it shows $\left<\sigma_{\rm ann}v_{\rm rel}\right>\lesssim 3\times10^{-31}$ for $m_\chi\approx10~{\rm GeV}$, based on previous data releases from the EHT collaboration \cite{Doeleman:2012zc,Akiyama:2015qta}. We estimate the EHT sensitivity as $\left<\sigma_{\rm ann}v_{\rm rel}\right>\lesssim 10^{-27}~{\rm cm^2/s}$ with $m_\chi\approx10~{\rm GeV}$ for an SIDM spike, which is comparable to the upper limit from the Milky Way; see the left panel of Fig.~\ref{fig:MW_M87}. Thus the EHT provides an interesting test of thermal SIDM models. Recently, the EHT collaboration observed the black hole shadow of M87 for the first time~\cite{Akiyama:2019eap}. It would be of interest to take their new results and further test the nature of dark matter, which we leave for future work.

\section{Conclusions}

Dark matter density spikes may form in the presence of a central black hole in galaxies. We have studied indirect detection constraints on dark matter annihilations after taking into account spikes predicted in SIDM and CDM models. For Draco, the upper limits on the cross section are not sensitive to the presence of an SIDM density spike ($\rho\propto r^{-7/4}$), if the black hole mass is reasonable for the system we consider. In contrast, the possibility of an existing intermediate black hole in Draco has been excluded for thermal relic CDM, as it predicts a steeper spike profile ($\rho\propto r^{-7/3}$). We further studied constraints from the Milky Way and M87, which host central supermassive black holes, and found that the upper limits on the annihilation cross section can be significantly weakened in SIDM. Observations from both galaxies exclude a thermal relic scenario for CDM for $s$-wave annihilations, but it's still allowed for SIDM. In addition, EHT observations of the M87 black hole can further probe the presence of an SIDM spike. In the future, we could study the distribution of SIDM particles near a black hole in the strong gravitational limit, which might be important for understanding growth of supermassive black holes in the early Universe, see., e.g.,~\cite{Balberg:2001qg,Feng:2020kxv,Choquette:2018lvq}.

\acknowledgments
We thank Ryan Keeley for useful discussion, Omid Sameie and Mauro Valli for providing data on Draco density profiles. We are grateful to the referee for critical comments on the first version of the manuscript. HBY was supported by the U.S. Department of Energy under Grant No. de-sc0008541, NASA 80NSSC20K0566, and the John Templeton Foundation \#61884.

\bibliography{Bibliography}

\end{document}